\documentclass[twoside]{article}
\usepackage{vmargin}
\usepackage{amsfonts}
\usepackage{graphics}
\usepackage{psfrag}
\newcommand{\beq}{\begin{equation}}
\newcommand{\eeq}{\end{equation}}
\newtheorem{Pn}{Proposition}
\setpapersize{USletter}
\setmarginsrb{39mm}{12mm}{28mm}{25mm}{12pt}{11mm}{0pt}{13mm}
		     
\begin{document}

\title{\Large\bf Separable Hilbert space in Loop Quantum Gravity}
\author{Winston Fairbairn\,$^{a}$\footnote{
fairbairn@cpt.univ-mrs.fr}\ , Carlo Rovelli\,$^{a,b}$\footnote{
rovelli@cpt.univ-mrs.fr}\\[1mm]
\normalsize {\itshape $^{a}\!$ Centre de Physique
Th\'eorique de Luminy, Case 907,
F-13288 Marseille, EU}\\ 
\normalsize{\em 
$^{b}\!$ Dipartimento di Fisica, Universit\`a ``La Sapienza",
P Moro 2, I-00185 Roma, EU}}
\date{{\small\today}}
\maketitle

\begin{abstract}
We study the separability of the state space of loop quantum gravity. 
In the standard construction, the kinematical Hilbert space of the
diffeomorphism-invariant states is nonseparable.  This is a
consequence of the fact that the knot-space of the equivalence classes
of graphs under diffeomorphisms is noncountable.  However, the
continuous moduli labeling these classes do not appear to affect the
physics of the theory.  We investigate the possibility that these
moduli could be only the consequence of a poor choice in the
fine-tuning of the mathematical setting.  We show that by simply
choosing a minor extension of the functional class of the classical
fields and coordinates, the moduli disappear, the knot classes become
countable, and the kinematical Hilbert space of loop quantum gravity
becomes separable.
\end{abstract}

%%%%%%%%%%%%%%%%%%%%%%%%%%%%%%%%%%%%%%%%%%%%%%%%%%%%%%%%%%%%%%%%%%%%%
\section{Introduction}
%%%%%%%%%%%%%%%%%%%%%%%%%%%%%%%%%%%%%%%%%%%%%%%%%%%%%%%%%%%%%%%%%%%%%

Loop quantum gravity (LQG) is a background-independent quantization of
classical general relativity which yields a discrete, combinatorial
picture of Planck-scale quantum geometry.\footnote{For an introduction
and references, see for instance \cite{Rovelli:2004mc}.} Quantum space
turns out to be described in terms of a basis of abstract spin-network
states, or $s$-knot states, labeled by discrete quantum numbers. 
However, the picture is not truly entirely discrete.  If the nodes of
the $s$-knots have sufficiently high valence (that is, when a
sufficiently high number of links meet), $s$-knots are labelled also
by certain {\em continuous} moduli parameters.  The existence of these
moduli was pointed out in \cite{Rovelli:1995ac} and their structure
studied in \cite{Grot:1996kj}.  Below we give an example of one of
these moduli explicitly.  These moduli are puzzling.  They are
virtually undetectable by the operators that represent physical
measurements, as well as by the hamiltonian operator that governs the
dynamics, and therefore they do not appear to play any significant
role in the theory.  Still, they spoil the discreetness of the picture
and they change the structure of the space of the
diffeomorphism-invariant states, ${\cal H}_{\rm diff}$, rather
drastically, making it nonseparable.\footnote{The space ${\cal H}_{\rm
diff}$ is denoted ${\cal K}_{\rm diff} $ in \cite{Rovelli:2004mc}.}

Nonseparability (absence of a countable basis) is generally regarded
as pathological in quantum field theory (QFT).  A classic discussion
on this issue is in \cite{Streater:1964qq}.  As pointed out in
\cite{Thiemann:2002nj}, the nonseparability of ${\cal H}_{\rm diff}$
is not necessarily unacceptable, because ${\cal H}_{\rm diff}$ is a
kinematical space that must still be reduced by the hamiltonian
constraint equation.  But it is nevertheless disturbing.  Do we have
to interpret it as an indication of a difficulty of the
background-independent loop quantization?

In this paper, motivated by the fact that the moduli do not appear to
have any physical significance, we consider the possibility that they
are indeed spurious.  We study the possibility that they are the
consequence of a poor choice in setting up the details of the
mathematical framework of the theory.  It is not unusual that a naive
way of setting up a QFT produces a nonseparable Hilbert space, which
is later cured, making it separable.  Indeed, recall that Fock space
itself, the prototypical QFT state space, was born precisely to cure
nonseparability.  A free scalar field can be decomposed into an
infinite set of oscillators $\varphi_{i},\ i=1,\ldots,\infty$.  If we
quantize each degree of freedom $\varphi_{i}$ as a standard quantum
harmonic oscillator, we are immediately led to a state space which is
the tensor product of an infinite number of separable Hilbert spaces
\beq {\cal H}= \bigotimes_{i=1,\ldots,\infty} H_{i}.  \eeq
A basis in ${\cal H}$ is given by an infinite sequence of nonnegative
integers $|n_{1},n_{2},\ldots, n_{i},\ldots, \rangle$, and such
infinite sequences are noncountable (for instance, the states with
$n_{i}< 10$ are put into one-to-one correspondence with the real
numbers $0\le x< 1$ by the decimal representation of $x$); hence
${\cal H}$ is nonseparable.  Fock found a way to circumvent the
problem by simply selecting the subspace $\cal F$ of ${\cal H}$
spanned by basis vectors where an arbitrary but finite number of
$n_{i}$ differ from zero.  It is $\cal F$, called today Fock space,
which is the appropriate state space for free QFT. Unlikely ${\cal
H}$, the Fock space $\cal F$ provides an {\em irreducible}
representation of the field algebra of the creation and annihilation
operators.  Thus, a straightforward and simple minded quantization
strategy leading to a nonseparable state space has been later
corrected to get rid of the nonseparability.  Can we do the same in
quantum gravity?  Are there physical reasons for doing the same in
quantum gravity?

In classical field theory, the choice of the functional class of
fields and coordinates is to a large extent just a matter of
convenience.  We can work with solutions of the field equations that
are smooth ($C^\infty$), or just twice differentiable ($C^2$),
sometimes distributional, or piecewise constant as in the lattice
approximation, or else, according to what is more convenient.  The
relation between the formal apparatus of field theory and reality is
only via the smearing of fields on finite regions of space, and
therefore we never empirically test the functional class of a physical
field.  In quantum gravity, the smooth category is generally chosen,
because it appears to be a natural and convenient setting.  In the
background-independent loop construction, the gauge invariance of
general relativity washes away virtually any remnant of the functional
class we started from, because the entire short-scale structure is
cancelled by the gauge transformations.  Virtually, but not entirely. 
Indeed, as we show below, the nodes of sufficiently high valence have
a surprising ``rigidity" under smooth transformation, and this
rigidity turns out to be the one responsible for the moduli. 
Therefore the non-separability of ${\cal H}_{\rm diff}$ is a bizarre
remnant of the initial choice of the smooth category.  It is therefore
natural to explore the possibility of using a slightly different
functional class of fields to start with.

Here we explore a minimal extension of this kind, where fields are
allowed isolated points of non-differentiability.  The gauge
invariance group of the theory becomes then a ``small" extension of
the diffeomorphism group, which we call {\em extended diffeomorphism}
and denote ${\it Diff}^*$.  As far as we can see, none of the physical
results, physical predictions or physical consequences of LQG studied
so far, are affected.  However, the knot classes, now formed by graphs
under ${\it Diff}^*$, turn out to be countable.  The continuous moduli
disappear and ${\cal H}_{\rm diff}$ becomes separable.  The theory
describes a quantum structure of spacetime that is genuinely
combinatorial at the Planck-scale.

These results can be taken as evidence that the moduli associated to
high-valence nodes are indeed spurious.  If we build LQG using the
setting described here, or a variant of the same, the moduli are not
anymore present, and the kinematical Hilbert space of the
diffeomorphism invariant states of LQG is separable.

In Section 2 we review the basic mathematical setting of LQG and the
origin of the nonseparability.  This article is mathematically
self-contained but for physical motivations and details see
\cite{Rovelli:2004mc}.  In Section 3 we introduce our main tool, ${\it
Diff}^*$, and show that the equivalence classes of graphs under ${\it
Diff}^*$ are countable and the resulting space ${\cal H}_{\rm diff}$
is separable.  In Section 4, we show that the modification introduced
has no effect on the geometrical operators of LQG. We discuss in
particular the area operator.  In Section 5 we summarize and conclude.

%%%%%%%%%%%%%%%%%%%%%%%%%%%%%%%%%%%%%%%%%%%%%%%%%%%%%%%%%%%%%%%%%%%%%
\section{Nonseparability}
%%%%%%%%%%%%%%%%%%%%%%%%%%%%%%%%%%%%%%%%%%%%%%%%%%%%%%%%%%%%%%%%%%%%%

%%%%%%%%%%%%%%%%%%%%%%%%%%%%%%%%%%%%%%%%%%%%%%%%%%%%%%%%%%%%%%%%%%%%%
\subsection{ ${\cal H}_{\rm diff}$, the space of the quantum-states of
physical-space}
%%%%%%%%%%%%%%%%%%%%%%%%%%%%%%%%%%%%%%%%%%%%%%%%%%%%%%%%%%%%%%%%%%%%%

We begin by briefly reviewing the basic mathematical setting of the
kinematics of LQG. Classical general relativity can be defined on a
four-dimensional differentiable manifold $M$ with topology $\Sigma\times
R$, where $\Sigma$ is a 3-dimensional differentiable manifold with a
fixed arbitrary topology.  The gravitational field can be taken to be
a smooth (that is, infinitely differentiable) pseudo-riemannian metric
tensor $g$ on $M$, satisfying the Einstein equations.  Call $\cal E$
the space of such fields.  The corresponding hamiltonian theory can be
defined on $\Sigma$.  The configuration variable of general relativity
can be defined as a smooth local connection one-form $A$ on a
$SU(2)$-principal bundle $\cal P$ over $\Sigma$.  Call $\mathcal{A}$ the
space of such connections.

We shall use extensively the notion of {\em graph}, whose relevance
for quantum gravity was first realized by Lewandowski.  Here a graph
$\Gamma $ is a finite collection of $L(\Gamma)$ smooth oriented
one-dimensional submanifolds of $\Sigma$, called links and noted $l$,
overlapping (if they do) only at endpoints, called nodes and denoted
$n$.  We say that the link $l$ ends at the node $n$ if $n$ is a
boundary point of $l$.  The {\em valence} of a node $n$ is the number
of links ending at $n$.  We call $\cal G$ the space of such graphs.

Quantum states are limit of sequences of cylindrical functions,
converging in the norm defined below.  A cylindrical function
$\Psi_{\Gamma,f}:\mathcal{A} \rightarrow \mathbb{C}$ is defined as
follows.  Let $U_l(A)\in SU(2)$ be the holonomy of the connection $A$
along $l$
\beq
U_l(A) \equiv \mathcal{P}\ exp \int_l A ,
\eeq
$\mathcal{P}$ denotes path ordering.  A graph $\Gamma
$ defines a map
$p_{\Gamma
}:{\mathcal{A}} \rightarrow [SU(2)]^{\times L(\Gamma
)}; \hspace{1mm}
A \rightarrow (U_l(A))$.  By composing this map with a complex valued
function $f$ on $[SU(2)]^{\times L(\Gamma
)}$, we obtain the cylindrical
function $\Psi_{\Gamma,f}=f \circ p_{\Gamma
}$, given by
\beq
\Psi_{\Gamma,f}(A) \equiv f(U_{l_1}(A),...,U_{l_{L(\Gamma
)}}(A)).
\eeq
Since it is always possible to write any two cylindrical functions in
terms of the same graph (a cylindrical function on a graph $\Gamma
$ can be
reexpressed on another graph $\Gamma
'$ that contains $\Gamma
$), we can define
a scalar product on the space of states using the Haar measure on
$SU(2)$
\beq
\langle \Psi_{(\Gamma , f)}|\Psi_{(\Gamma , f')}\rangle
\equiv \int
dU_{1}...dU_{L(\Gamma
)}
\hspace{2mm} \overline{f(U_{1},...,U_{L(\Gamma
)})} \hspace{1mm}
f'(U_{1},...,U_{L(\Gamma
)}) .
\label{sp}
\eeq
The kinematical Hilbert space $\cal K$ of LQG is defined as the
completion in the Hilbert norm (\ref{sp}) of the space of the
cylindrical functions.

Local $SU(2)$ gauge transformations act naturally on this space and
the invariant states form a proper subspace ${\cal K}_{0}$.  An
orthonormal basis in ${\cal K}_{0}$ can be obtained using the
spin-network states \cite{Rovelli:1995ac, Baez:1995md}.  Consider a
graph $\Gamma $ and color each link $l$ with a unitary irreducible
representations $j_l$ of $SU(2)$.  At each node $n$, fix a basis in
the space of the invariant elements (the ``intertwiners") in the
tensor product of the representation spaces associated to the links
that end at $n$, and color the node with an intertwiner $i_{n}$ of
this basis (see \cite{Rovelli:2004mc} for details).  The triple
$S=(\Gamma ,j_{l},i_{n})$ is called a spin network.  Let the function
$f_{S}$ be defined by the representation matrices for each link $l$ in
the representation $j_{l}$, contracted with the invariant tensors at
each node.  The state $\Psi_S[A] \equiv\Psi_{\Gamma ,f_S}[A] \equiv
\langle A | S \rangle$ is a spin network state.  Varying the graph
$\Gamma $ in $\cal G$, the irreducible representations $j_{l}$ and the
intertwiners $i_{n}$ in the chosen bases, we obtain an orthonormal
basis in ${\cal K}_{0}$.

Finite linear combinations of spin-network states form a dense
subspace $\cal S$ in ${\cal K}_{0}$.  Sequences of states that
converge weakly on $\cal S$, form the dual ${\cal S}'$.  The Gelfand
triple $({\cal S}\subset {\cal K}_{0}\subset {\cal S}')$ describes the
space of the $SU(2)$ invariant states of the theory. Now, ${\cal K}_{0}$
carries a natural unitary representation $\Psi[A] \rightarrow U_{\phi}
\Psi[A] = \Psi[\phi^{-1}A]$ of the group ${\it Diff}_{\Sigma}$ of the
diffeomorphisms of $\Sigma$
\beq
\phi: \Sigma \to \Sigma\ ,
\eeq
which extends naturally to ${\cal S}'$ by duality (we have indicated
with $\phi$ also the pull back action generated by the diffeomorphism
$\phi$ on forms).  The diffeomorphism invariant states form a linear
subspace ${\cal H}_{\rm diff} $ of ${\cal S}'$.  This
space describes the diff-invariant quantum states of
the gravitational field; that is, the quantum states of physical
space.

A map $P_{{\rm diff}}: {\cal S} \rightarrow {\cal S}'$ is naturally
defined by
\beq
(P_{{\rm diff
}}\Psi)(\Psi')= \sum_{\Psi''=U_{\phi} \Psi} \langle
\Psi'' , \Psi' \rangle, 
\label{P}
\eeq
where the sum is over al distinct states $\Psi''$ that are equal to
$U_{\phi} \Psi$ for some $\phi\in {\it Diff}$.  This sum converges and
is well defined.  The state $P_{{\rm diff}}\Psi\in{\cal S}' $ is
diff-invariant because clearly states related by diffeomorphisms are
projected by $P_{{\rm diff }}$ on the same element of
$\mathcal{H}_{{\rm diff }}$
\beq
P_{{\rm diff }} \Psi_S = P_{{\rm diff }}(U_{\phi} \Psi_S).
\eeq
Furthermore, the states of this form span ${\cal H}_{\rm diff}$.  The
linear space $\mathcal{H}_{{\rm diff}}$ is naturally equipped with a
Hilbert space structure by the scalar product
\beq
\langle P_{{\rm diff }}\Psi_S , P_{{\rm diff }}\Psi_{S'}
\rangle_{\mathcal{H}_{{\rm diff }}} \equiv (P_{{\rm diff
}}\Psi_S)(\Psi_{S'}).
\eeq
Equivalently, the Hilbert space $\mathcal{H}_{{\rm diff }}$ can be
defined by the bilinear form
\beq
\langle \Psi , \Psi' \rangle_{\mathcal{H}_{{\rm diff }}} \equiv
\langle \Psi | P_{{\rm diff }} | \Psi' \rangle \equiv
\sum_{\Psi''=U_{\phi} \Psi} \langle \Psi'' , \Psi' \rangle.
\eeq

We can unravel the structure of $\mathcal{H}_{{\rm diff }}$ by
studying the action of a diffeomorphism on a spin network state.
Since the holonomy transforms as
\beq 
 U_l(\phi^{-1} A) = U_{\phi \circ l}(A),
\eeq
shifting $A$ with $\phi\in {\it Diff}(\Sigma)$ is equivalent to shifting 
the
curve $l$.  Consequently, the (representation of the) spatial
diffeomorphism group acts directly on the spin network of a
spin-network state:
\beq
U_{\phi} | S \rangle = | \phi \circ S \rangle .
\eeq
Since $|\phi \circ S \rangle$ may be formed by invariant tensors that are not 
in the intertwiner basis chosen at the nodes, $|\phi \circ S \rangle$ may fail
to be a basis state even if $|S\rangle$ is; but as the spaces of
intertwiners are finite dimensional, it will be a finite linear
combination of basis states, all having the same graph
$\Gamma'=\phi \circ \Gamma$.  In particular, given a spin network state,
there is a finite group of transformations $g_{k}$, $k=1,\ldots,K$, in
the (tensor over the nodes of the) spaces of the intertwiners, that can
be obtained by a diffeomorphism mapping the graph into itself.  We can
therefore write
\beq
\langle S | P_{{\rm diff
}} | S'\rangle =
\left \{ \begin{array}{cc} 0 & \mbox{if}\ \hspace{1mm}
\Gamma
' \neq \phi \circ \Gamma
, \\
\sum_k \langle S | g_k | S'\rangle      & \mbox{if}\ \hspace{1mm}
\Gamma
'  = \phi \circ \Gamma
. \\
\end{array} \right.
\label{gk}
\eeq
{} From the first line, we see that two spin networks $S$ and $S'$ define
orthogonal states in $\mathcal{H}_{{\rm diff }}$ if the corresponding
graphs $\Gamma$ and $\Gamma'$ belong to different equivalence classes under
diffeomorphism transformations.  We call these equivalence classes
``diff-knot classes" and indicate them as $K_{\rm d}$, where ``d" is
for diffeomorphism.  The basis states in $\mathcal{H}_{{\rm diff }}$
are therefore firstly labelled by a diff-knot class $K_{\rm d}$.  Call
$\mathcal{H}_{K_{\rm d}}$ the subspace of $\mathcal{H}_{{\rm diff }}$
spanned by the basis states labelled by the knot class $K_{\rm d}$.
The states in $\mathcal{H}_{K_{\rm d}}$ are then only distinguished by
the coloring of links and nodes.  Hence $\mathcal{H}_{K_{\rm d}}$ is
separable. These colorings are however not necessarily orthogonal because
of the nontrivial action of the discrete group.  A diagonalization of
the matrix of the bilinear form in the second line of expression
(\ref{gk}) yields states $|s \rangle = |K_{\rm d} , c \rangle$, where
the discrete label $c$ depends only on colorings.  The states $|s
\rangle = |K_{\rm d} , c \rangle$ are called $s$-knot states.

If $|s \rangle = P_{{\rm diff }}|S \rangle$, the state $|s \rangle$
represents the diffeomorphism equivalence class to which the spin
network $S$ belongs.  In going from from the spin network state $|S
\rangle$ to the $s$-knot state $|s \rangle$, we preserve the
information in $|S \rangle$ except for it's location in $\Sigma$. 
This is the quantum analog to the fact that physically distinguishable
solutions of the classical Einstein equations are not fields, but
equivalence classes of fields under diffeomorphisms.  It reflects the
core of the conceptual revolution of general relativity: spatial
localization concerns only the {\em relative} location of the
dynamical fields, and not their location in a background space. 
Accordingly, the $s$-knot states are not quantum excitations {\em
in\/} space, they are quantum excitation {\em of\/} space. An
$s$-knot does not reside ``somewhere": the $s$-knot itself defines the
``where".

However, a remnant of the background structure oddly remains, when a
node has sufficiently high valence, as we show in the next section.

%%%%%%%%%%%%%%%%%%%%%%%%%%%%%%%%%%%%%%%%%%%%%%%%%%%%%%%%%%%%%%%%%%%%
\subsection{Moduli space structure}
%%%%%%%%%%%%%%%%%%%%%%%%%%%%%%%%%%%%%%%%%%%%%%%%%%%%%%%%%%%%%%%%%%%%

We have seen above that quantum states of the gravitational field are
labelled by diff-knot classes.  Following \cite{Grot:1996kj}, we now
study the structure of these classes and the way this affects the
structure of $\mathcal{H}_{{\rm diff}}$.  The key point that we
underline below is the fact that diff-knots are not countable, hence
$\mathcal{H}_{{\rm diff}}$ is nonseparable.

To begin with, consider usual knots, namely ones without
intersections.  These can be defined in two equivalent manners.  Let
$\cal L$ be the space of loops without intersections, namely the space
of smooth embeddings of $S_{1}$ into $\Sigma$.  We denote loops in
$\cal L$ as $\alpha, \beta, \ldots$ .  Consider {\em two} equivalence
relations on this space.  First (see for instance \cite{GB}) we say
that $\alpha$ and $\beta$ are iso-equivalent, and write $\alpha
\sim_{\rm i} \beta$, if there is a {\em continuous} ambient isotopy
relating the two; that is, a one parameter family of {\em homeomorphisms} 
$h_t:\Sigma \rightarrow \Sigma, \hspace{1mm} t \in [0,1]$, such that $h_0$
is the identity map on $\Sigma$ while $h_1$ maps $\alpha$ to $\beta$.
We call the equivalence classes of loops in $\cal L$ under this equivalence
relation ``iso-knots".  Next, (see for instance \cite{BaezMunian}) we
say that $\alpha$ and $\beta$ are diff-equivalent, and write $\alpha
\sim_{\rm d} \beta$ if there exists a {\em diffeomorphism} $\phi$ of
$\Sigma$ in the connected component of the identity, such that $\alpha
= \phi\circ\beta$ (or, equivalently, if there exists a {\em smooth} ambient 
isotopy relating the two).  We denote the corresponding equivalence classes in
$\mathcal{L}$ as ``diff-knots".  A classic result of knot theory states
that two loops are diff-equivalent if and only if they are
iso-equivalent.  Hence diffeomorphism equivalence is the same as
isotopy, as far as loops without intersections are concerned, and a
diff-knot is also an iso-knot.\footnote{This is also equivalent to the
existence of a {\em smooth homotopy} between $\alpha$ and $\beta$,
that is, a smooth one-parameter family of embeddings $\alpha_{t}$, $t
\in [0,1]$, such that $\alpha_{0}=\alpha$ and $\alpha_1=\beta$. 
However, this is {\em not} equivalent to to the existence of a {\em
continuous homotopy} between $\alpha$ and $\beta$, since, perhaps
surprisingly, any two knots can be related by continuous homotopy. 
See for instance Ref \cite{Liv} page 14 or Ref \cite{GB}, exercise
3.5.4 page 53.}

However, this result does not extend to the case in which
intersections, or nodes, are present; something peculiar happens at
the intersection points.  Let $\cal G$ be the space of the graphs
defined in the previous section, of which $\cal L$ is a subset. 
Define isotopy equivalence and diff-equivalence between graphs
precisely as we did for loops.  Let's indicate iso-knots by $K_{\rm
i}$ and the space of the iso-knots as $\mathcal{K}_{\rm i}$. 
Similarly, let's indicate diff-knots by $K_{\rm d}$ and the space of
the diff-knots as $\mathcal{K}_{\rm d}$.  Thus
\beq
\mathcal{K}_{\rm i} = \frac{\mathcal{G}}{\sim_{\rm i}}, \hspace{5mm}
\mathcal{K}_{\rm d} = \frac{\mathcal{G}}{\sim_{\rm d}} .
\eeq
It is still easy to see that diff-equivalence implies iso-equivalence,
because a diffeomorphism in the connected component of the identity is indeed
connected to the identity by a smooth one parameter family of diffeomorphisms
which generates the isotopy. But now
the converse is not true: there are iso-equivalent graphs that are not
diff-equivalent.  Furthermore, while iso-knots are countable,
diff-knots are not: they are distinguished by moduli. (In general,
a ``modulus" is a continuous parameter labelling equivalence classes.)

The roots of this fact, which at first might seem surprising, can be
illustrated using a simple example presented by Arnold in his book on
catastrophe theory \cite{Arnold:1984kk}.  Let $L_{n}$ be the set of
all $n$-tuples of lines in the plane, passing through the origin.  Say
that two $n$-tuples are equivalent if they can be mapped into each
other by a linear transformation of the plane, and let $T_{n}$ be the
space of the equivalence classes of $n$-tuples under this equivalence
relation.  The spaces $T_{1}, T_{2}$ and $T_{3}$ are discrete, but
$T_{4}$ is a one-dimensional space.  Indeed the group of linear
transformation of the plane, $GL(2)$, is four dimensional, but
it does not act effectively on $T_n$ because the 
dilatations do not affect lines through the origin, hence we can
consider just its $SL(2)$ subgroup, which is three dimensional.  But
$L_{4}$ is four dimensional, and the dimension of $T_{4}=L_{4}/SL(2)$
is $4-3=1$.  More explicitly, if $\phi_{1}, \ldots, \phi_{4}$ are the
angles formed by the lines with a given axis, then it is easy to see
that
\beq
\lambda(\phi_{1}, \ldots, \phi_{4}) =
\frac{\sin(\phi_{1}-\phi_{3})\ \sin(\phi_{2}-\phi_{4})}
{\sin(\phi_{1}-\phi_{4})\ \sin(\phi_{2}-\phi_{3})}
\label{lambda}
\eeq
is invariant under linear transformations.  Indeed, let $\vec v_{1},
\ldots, \vec v_{4}$ be four vectors of arbitrary length parallel to
the four lines.  The ratio
\beq
\lambda(\vec v_{1}, \ldots, \vec v_{4}) = 
\frac{({\vec v_{1}\times\vec v_{3})\ (\vec v_{2}
\times\vec v_{4}})}{({\vec
v_{1}\times\vec v_{4})\ (\vec v_{2}\times\vec v_{3}})}, 
     \label{lambda1}
\eeq
where $\vec v_{1}\times\vec v_{2}=\det(\vec v_1,\vec v_2)$ is the 2d
vector product, is invariant under linear transformations because the
vector product transforms with the determinant of the linear
transformation.  On the other hand, the ratio does not depend on the
length of the vectors, and is equal to (\ref{lambda}).  Therefore
$\lambda$ is a continuous modulus that distinguishes $GL(2)$ equivalence
classes in $L_{4}$, and $T_4$ is a continuous space.

The same happens in three dimensions.  Here $GL(3)$ has nine
dimensions, of which only eight affect the $n$-tuplets of lines
through the origin; five lines through the origin are determined by
ten angles; hence in 3d $T_{n}=L_{n}/GL(3)$ is only discrete for
$n<5$.  If $\vec v_{1}, \ldots, \vec v_{5}$ are five vectors of
arbitrary length parallel to the five lines, the ratio
\beq
\lambda(\vec v_{1}, \ldots, \vec v_{5}) =
\frac{\vec v_{1}\cdot({\vec v_{2}\times\vec v_{3})\ \
\vec v_{1}\cdot(\vec v_{4}\times\vec v_{5}})}
 {\vec v_{1}\cdot({\vec v_{2}\times\vec v_{5})\  \
\vec v_{1}\cdot(\vec v_{4}\times\vec v_{3}})}, 
    \label{lambda2}
\eeq
where $\vec v_{1}\cdot(\vec v_{2}\times\vec v_{3})= \det(\vec v_1,\vec
v_2,\vec v_3)$ is the triple product, is invariant under linear
transformation.  It classifies quintuplets of straight lines into 
equivalence classes.  The invariant (\ref{lambda1}) is well known in
projective geometry since the nineteenth century, as the {\em
cross-ratio}.  It distinguishes orbits of quadruplets of points
generated by the action of the projective group on the real projective
line.  The generalization of the cross-ratio projective invariant to
higher dimensions has been studied by Hilbert.

Let us return to nodes.  Consider for simplicity a graph $\Gamma$ with
a single node $n$.  Say that the node is five-valent.  Let $\Gamma'$
be a node iso-equivalent to $\Gamma$.  The graph $\Gamma'$ will have a
single five-valent node $n'$ as well.  Say, for simplicity, that $n'$
is located at the same point $p$ as $n$.  Can we always find a
diffeomorphism $\phi$ that sends $\Gamma'$ into $\Gamma$?  A condition
on $\phi$ is that $\phi(p)=p$.  The tangent map $\phi_*:T_{p} \to
T_{p}$ defines a linear action on the tangent space $T_{p}$ of $\Sigma$
at $p$.  The tangents to the five links of $\Gamma$ at $p$ determine a
quintuplet of straight lines in $T_{p}$.  Similarly, the tangents to
the links of $\Gamma'$ determine another quintuplet of straight lines
in $T_{p}$.  A condition on $\phi$ is then that the linear
transformation $\phi_*$ maps the first quintuplet of lines into the
second.  As observed above, in general such a linear transformation
does not exist.  Equivalence classes under linear transformations of
quintuplets of lines are distinguished by continuous moduli.  If $\vec
v_i(\Gamma,n),\ i=1,\ldots,5$ are the tangents to the five links of
$\Gamma$ at the node $n$, in an arbitrary parametrization and an
arbitrary coordinate system, the quantity
\beq
\lambda(\Gamma,n) =
\lambda(\vec v_{1}(\Gamma,n), \ldots, \vec v_{5}(\Gamma,n)),
\label{lambda4}
\eeq
where the function on the right hand side is defined in
(\ref{lambda2}), does not depend neither on the parametrization nor on
the coordinates chosen, and is invariant under diffeomorphisms.  Hence
if $\lambda(\Gamma
,n)\ne\lambda(\Gamma
',n')$, the two iso-equivalent graphs
$\Gamma
$ and $\Gamma
'$ are not diff-equivalent.  The function (\ref{lambda4})
is an example of the continuous moduli that distinguish diff-knots.

In general, a single iso-knot $K_{\rm i}$ is therefore formed by a
continuous set of diff-knots $K_{\rm d}$; diff-knots are labelled by
moduli at the intersections.  These moduli have a structure far richer
than Hilbert's projective invariants.  Indeed, it is not only the
tangent structure that may distinguish iso-equivalent nodes, but also,
for sufficiently high valence, the higher derivatives of the links at
the intersection.  For instance, assume that the tangents of the links
at the node of two graphs are the same, but their curvatures differ. 
Under a diffeomorphism $\phi$, the transformation of these curvatures
is determined by the second derivatives of $\phi$ at the node, but
these are finite in number as well (they are 18) and therefore for a
sufficiently high valence they are not sufficient, in general, to map
all curvatures of one graph into the curvatures of the second.  We
refer to \cite{Grot:1996kj} for a detailed analysis of the structure
of the moduli.

We have seen in the previous section that the diff-knots $K_{\rm d}$
label a basis in ${\cal H}_{\rm diff}$.  Since they are not countable,
the kinematical state space ${\cal H}_{\rm diff}$ space admits a
continuous basis and therefore is nonseparable.  Thus the root of the
nonseparability of the state space of LQG is the ``rigidity" of the
diffeomorphisms at isolated points: the differential structure of the
underlying manifold is rigid in the sense that it produces the linear
structure of the tangent spaces $T_{p}$, which leaves quantities such
as the cross-ratio (\ref{lambda}) invariant.

%%%%%%%%%%%%%%%%%%%%%%%%%%%%%%%%%%%%%%%%%%%%%%%%%%%%%%%%%%%%%%%%%%%%
\subsection{Discussion}
%%%%%%%%%%%%%%%%%%%%%%%%%%%%%%%%%%%%%%%%%%%%%%%%%%%%%%%%%%%%%%%%%%%%

The nonseparability of the kinematical Hilbert space ${\cal H}_{\rm
diff}$ is disturbing for several reasons.

First, the background independence of general relativity implies that
the localization of the dynamical fields on the coordinate manifold
has no physical meaning: only relative localization of dynamical
objects with respect to one another is physically significant.  In the
classical theory, background independence is implemented by the fact
that diffeomorphisms turn out to be gauges.  When implementing gauge
invariance in the quantum theory, the localization of the spin network
in the manifold is washed away by the gauge transformation, and only
the discrete combinatorial relations remain -- but not completely so. 
The moduli distinguishing diff-knots are a remnant of the localization
of the spin network in the coordinate manifold.  It is difficult to
reconcile the presence of this remnant with the physical principle,
underlying general relativity, that wants the localization on the
coordinate manifold to be physically irrelevant.

Second, recall that loop states and spin-network states form a good
basis in lattice Yang-Mills theory, but in continuous Yang-Mills
theory they are ``too singular" and ``too many", because of
their continuous dependence on position.  In gravity, the continuous
dependence of position is gauged away by diffeomorphism invariance,
dramatically reducing the size of the resulting state space
\cite{Rovelli:1988,Rovelli:1995}.  This is the reason for
which the loop basis becomes viable thanks to background independence,
and therefore the rationale underlying the background-independent loop
quantization.  It is quite puzzling that this dramatic reduction of
the state space fails to be complete because of the moduli. 

Third, if we accept the formalization of generally covariant quantum
theory described in \cite{Rovelli:2004mc}, the Hilbert space ${\cal
H}_{\rm diff}$ describes distinguishable quantum states.  A realistic
space of distinguishable quantum states should be described by a
separable Hilbert space \cite{Streater:1964qq}.

Now, if the continuous moduli had a physical meaning, they should
affect measurements, or affect the dynamics.  Two states that differ
only by different values of their moduli should in principle be
distinguishable by means of physical measurements.  However, as
mentioned in the introduction, this does not appear to be the case. 
The only effect of the moduli appears to be to vastly enlarging the
kinematical Hilbert space, with no visible effect on the physics of
the theory.  One is therefore naturally led to the idea that these
moduli may be spurious.

Above we have observed that the moduli are a consequence of the
incapacity of diffeomorphisms to act at vertices in a way sufficiently
general to map iso-equivalent graphs into each other.  Diffeomorphisms
are ``rigid" at vertices.  If gauge transformations included maps
$\phi: \Sigma\to \Sigma$ less rigid at nodes (homeomorphisms, for
instance), the continuous moduli would disappear: states distinguished
by different moduli would become gauge equivalent.  So, it is natural
to ask: what is it that forces gauge transformations $\phi:
\Sigma\to \Sigma$ implementing background-independence, to be
diffeomorphisms?  Namely to be smooth?

To a large extent, the answer is: just the conventional way we set up
the theory.  It is simple and convenient to use smooth fields on a
smooth manifold, and stay in the smooth category.  The moduli are a
consequence of this choice, which might have little do to with
physics.  It is therefore natural to investigate the possibility of a
different mathematical starting point, that would not affect classical
physics, but would free the quantum theory from the moduli.

A very natural setting one may consider is the piecewise smooth
category.  Another possibility, investigated by Jos\`e Zapata
\cite{Zapata} is to start from a piecewise linear manifold.  With
these choices, the moduli disappear and $\mathcal{H}_{{\rm diff}}$
becomes separable.  In this paper we investigate a minimal choice: we
consider fields that are {\em everywhere continuous and smooth
everywhere except possibly at a finite number of points}.  We call
these fields ``almost smooth".  This minor enlargement of the space of
the fields has practically no effect on the classical theory, nor on
the physical results of LQG. But it gets rid of moduli and
nonseparability.

%%%%%%%%%%%%%%%%%%%%%%%%%%%%%%%%%%%%%%%%%%%%%%%%%%%%%%%%%%%%%%%%%%%%%
\section{Extended diffeomorphism group}
%%%%%%%%%%%%%%%%%%%%%%%%%%%%%%%%%%%%%%%%%%%%%%%%%%%%%%%%%%%%%%%%%%%%%

We now define a modified theory, where the gauge group is an extension
of the diffeomorphism group.  We show that the knot classes defined by
the equivalence relation determined by this extended gauge group are
countable, and lead to a separable ${\cal H}_{\rm diff}$.

%%%%%%%%%%%%%%%%%%%%%%%%%%%%%%%%%%%%%%%%%%%%%%%%%%%%%%%%%%%%%%%%%%%%%
\subsection{Almost smooth physical fields}
%%%%%%%%%%%%%%%%%%%%%%%%%%%%%%%%%%%%%%%%%%%%%%%%%%%%%%%%%%%%%%%%%%%%

Consider a four-dimensional differentiable manifold $M$ with topology
$\Sigma\times R$, as before.  However, we now allow the gravitational
field $g$ to be almost smooth, as defined in the previous section, 
that is: $g$ is a continuous field which is smooth everywhere except
possibly at a finite number of points, which we call the singular
points of $g$.  Any such $g$ can be seen as a (pointwise) limit of a
sequence of smooth fields.  We say that $g$ is a solution of the
Einstein equations if it is the limit of a sequence of a sequence of
smooth solutions of the Einstein equations.  Call ${\cal E}^{*}$ the
space of such fields.

Let now $\phi$ be an invertible map from $M$ to $M$ such that
$\phi$ and $\phi^{-1}$ are continuous and are infinitely differentiable
everywhere except possibly at a finite number of points.  The space of
these maps form a group under composition, because the composition of
two homeomorphisms that are smooth except at a finite number of
singular points is clearly an homeomorphisms which is smooth except at
a finite number of singular points.  We call this group the {\em
extended diffeomorphism group} and we denote it as ${\it Diff}_{M}^*$.  
It
is clear that if $g\in{\cal E}^*$ then $(\phi g)\in{\cal E}^*$ for any
$\phi \in {\it Diff}_{M}^*$.  Hence ${\it Diff}_{M}^*$ is a gauge group 
for the theory.

In the Hamiltonian theory, we can now take almost smooth connections
$A$ on $\Sigma$.  Notice that the holonomy of an almost smooth connection
on a link $l$ is well defined, because it is the product of the
holonomies on the portions in which $l$ is partitioned by the eventual
singular points of $A$.  We can thus define cylindrical functions,
$\cal K$ and ${\cal K}_{0}$ as before.  However, the gauge group ${\it
Diff}$ is now replaced by the gauge group ${\it Diff}^*$, formed by
the homeomorphisms of $\Sigma$ that are almost smooth (with their
inverse).  The group ${\it Diff}$ considered above is a dense subgroup
of ${\it Diff}^*$ \footnote{${\it Diff}^*$ can be given a topological
group structure as a subgroup of the homeomorphism group of
$\Sigma$.  The question of whether it can be given a Lie group structure
is more difficult.}.  Notice that ${\it Diff}^*$ has a well defined
action on the space of the graphs $\cal G$.  Unlike ${\it Diff}$,
${\it Diff}^*$ does not preserve the number of nodes of a graph,
because a singular point of $\phi$ may break a link into two links and
create a bivalent node.

The construction ${\cal H}_{\rm diff}$ is the same as before, with the
only difference that $\phi$ in (\ref{gk}) is now in ${\it Diff}^*$.  
Hence
${\cal H}_{\rm diff}$ is now spanned by a basis of states $|s \rangle
= |K_{\rm d^*}, c \rangle$ where $c$ is a discrete quantum number as
before, but $K_{\rm d^*}$ is an element of $\mathcal{K}_{\rm d^*}$,
namely an equivalence class of graphs under {\em extended}
diffeomorphisms
\beq 
\mathcal{K}_{\rm d^*} = \frac{\mathcal{G}}{\sim_{\rm d^*}}, 
\eeq
where $\Gamma
\sim_{\rm d^*}\Gamma'$ if there is a $\phi\in {\it Diff}^*$ such that
$\Gamma'=\phi \circ \Gamma$.  We denote the elements $K_{\rm d^*}$ of
$\mathcal{K}_{\rm d^*}$ as diff$^*$-knots.

%%%%%%%%%%%%%%%%%%%%%%%%%%%%%%%%%%%%%%%%%%%%%%%%%%%%%%%%%%%%%%%%%%%%%%%
\subsection{ Diff$^*$-knots are countable}
%%%%%%%%%%%%%%%%%%%%%%%%%%%%%%%%%%%%%%%%%%%%%%%%%%%%%%%%%%%%%%%%%%%%%%%

We now prove that diff$^*$-knots are countable.  Two diff-equivalent
graphs are also diff$^*$-equivalent, because ${\it Diff}$ is a
subgroup of ${\it Diff}^*$.  Therefore diff$^*$-knots are equivalence
classes of diff-knots.  To prove that diff$^*$-knots are countable it
is sufficient to prove that any two diff-knots distinguished
by a continuous parameter are diff$^*$-equivalent.  For this, it is
sufficient to consider two iso-equivalent but diff-inequivalent graphs
$\Gamma$ and $\Gamma'$.  Our strategy will be to explicitly build an extended
diffeomorphism mapping $\Gamma$ into $\Gamma'$  (see Figure 1).  Since
iso-knots are countable, this will be sufficient to show that
$\mathcal{K}_{d^*}$ is countable.
 
\begin{figure}[t]
\begin{center}
\psfrag{a}{$\phi$}
\psfrag{b}{$\Gamma$}
\psfrag{c}{$\Gamma'$}
\includegraphics{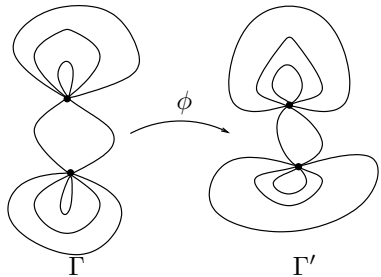}
\caption{Two graphs living in the same iso-knot but not in the same
diff-knot.}
\end{center}
\end{figure}

Choose an arbitrary smooth metric $g$ on $\Sigma$.  Consider a node
$n$ of $\Gamma$, say with valence $v$. Consider the open ball $B_{n}$ 
of fixed radius $\epsilon(n)$ about $n$. 
Let $S_{r}$ be the spheres of radius $r(n)$, $0<r(n)<\epsilon(n)$, 
centered at the same point $n$.  Introduce (nonmetrical) angular
coordinates $(\theta,\phi)_{r}$ on each sphere $S_{r}$ of radius $r$,
smooth in $r$.  By choosing $\epsilon$ sufficiently small, each link
$l$ ending at $n$ will intersect each sphere only once, say in the
point with coordinates $(\theta_{l},\phi_{l})_{r}$.  (No fold-backs,
see Figure 2.)  Now, we can always choose the coordinates
$(\theta,\phi)_{r}$ in such a way that the coordinates
$(\theta_{l},\phi_{l})_{r}$ of the link $l$ are independent from $r$,
and are equal to $v$ arbitrarily chosen values $(\theta_{l},\phi_{l})$. 
This construction can be repeated for the node $n'$ of $\Gamma'$ which is
isotopically associated to the node $n$.  For the coordinates on the
ball $B_{n'}$ around the node $n'$ we use the notation
$(r',\theta',\phi')$ and we choose $(\theta'_{l'}=\theta_{l},
\phi_{l'}'=\phi_{l})$, namely equal angular coordinates for
corresponding links.  We call $\Gamma_{n}$ the restriction of $\Gamma$ to
$B_{n}$ and $\Gamma'_{n}$ the restriction of $\Gamma'$ to $B_{n'}$.

Consider now the space $\Sigma \backslash B_{n}$ obtained by removing all 
balls
$B_{n}$, for all nodes $n$, from $\Sigma$, and the space 
$\Sigma\backslash B'_{n}$
obtained removing all balls $B_{n'}$.  Let $\tilde\Gamma$ be the
restriction of $\Gamma$ to $\Sigma\backslash B_{n}$ and $\tilde\Gamma'$ 
be the
restriction of $\Gamma'$ to $\Sigma\backslash B_{n'}$.  Under the 
hypothesis we made
that $\Gamma$ and $\Gamma'$ are iso-equivalent, a smooth invertible map
\begin{eqnarray} 
\tilde\phi: \Sigma\backslash B_{n} &\to& \Sigma\backslash B'_{n}, 
\nonumber \\
\tilde\Gamma\ \ \ &\mapsto& \ \ \ \tilde\Gamma' 
\end{eqnarray}
exists, because we are here in the simpler case of loops without
intersections (on a space with boundaries), where standard knot-theory
results apply. The failure of iso-equivalence to
yield diffeo-equivalence regards {\em only} the neighborhoods of the
nodes.

To prove that  $\Gamma$ and $\Gamma'$ are diff$^*$-equivalent, we have 
therefore just to construct maps 
\begin{eqnarray}
\phi_{n}: B_{n} &\to& B'_{n}, \nonumber \\
\Gamma_{n} &\mapsto& \Gamma_{n}' 
\end{eqnarray}
such that $\tilde\phi$ and $\phi_{n}$, taken together, give an almost 
smooth map $\phi: \Sigma\to\Sigma$.  Let $\phi_{n}$ be given simply by 
\begin{eqnarray}
\phi_{n}: B_{n} &\to& B'_{n}, \nonumber \\
(r,\theta,\phi) \ &\mapsto& \ (r'=r,\theta'=\theta,\phi'=\phi). 
\end{eqnarray}
$\tilde\phi$ can be chosen so that at the boundaries of the balls
$\phi$ is smooth.  Hence $\phi$ is smooth for all $r>0$.  The map
$\phi_{n}$ can immediately be continued to $r=0$, yielding by
continuity $\phi_{n}(n)=n'$.  But there is no reason for this
continuation to be smooth, and in fact, in general it will not be. 
Hence $\phi$ is not in ${\it Diff}$.  But it is in ${\it Diff}^*$,
because it is continuous, invertible and smooth everywhere except at
the nodes, which are finite in number.  Therefore iso-equivalent
graphs $\Gamma$ and $\Gamma'$ are diff$^*$-equivalent.  Therefore
\begin{Pn}
The space of the diff$^*$-knots ${\cal K}_{d^*}$ is countable.
\end{Pn}
It follows immediately that
\begin{Pn}
If $\phi\in {\it {\it Diff}}^*$, the space ${\cal H}_{\rm diff}$
defined by the bilinear form (\ref{gk}) is separable.
\end{Pn}
Therefore we have shown that a minor extension of the functional
space of the fields considered eliminates the continuous moduli and
the nonseparability of the kinematical state space of LQG.

\begin{figure}[t]
\begin{center}
\psfrag{a}{$B_n$}
\includegraphics{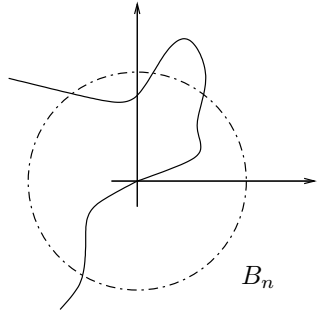}
\caption{A Link presenting fold-backs.}
\end{center}
\end{figure}

%%%%%%%%%%%%%%%%%%%%%%%%%%%%%%%%%%%%%%%%%%%%%%%%%%%%%%%%%%%%%%%%%%%%%
\section{Operators}
%%%%%%%%%%%%%%%%%%%%%%%%%%%%%%%%%%%%%%%%%%%%%%%%%%%%%%%%%%%%%%%%%%%%%

In this section we discuss some consequences of the extension of ${\it
Diff}$ to ${\it Diff}^*$ and clarify some apparent difficulties that
this extension raises.

%%%%%%%%%%%%%%%%%%%%%%%%%%%%%%%%%%%%%%%%%%%%%%%%%%%%%%
\subsection{Conical singularities and area operator}
%%%%%%%%%%%%%%%%%%%%%%%%%%%%%%%%%%%%%%%%%%%%%%%%%%%%%%

If the gauge group is ${\it Diff}^*$, a smooth 2-dimensional surface
is gauge equivalent to a ``singular" surface, that is a surface with
conical singularities.  The area operator $A({\cal S})$ of LQG has
been defined for smooth surfaces $\cal S$.  Is it well defined also
for a singular surface ${\cal S}$?  Naively, one may think that
$A({\cal S})$ is ill defined for a singular $\cal S$, for the
following reason.  Consider a two-dimensional surface ${\cal S}$
embedded in $\Sigma$.  Let $x=(x^a), a=1,2,3$ be coordinates on $\Sigma$ 
and
$u=(u^m), m=1,2$ coordinates on ${\cal S}$; the embedding is given by
the functions $x^a(u)$.  Let $g_{ab}(x)$ be the 3d metric, namely the
gravitational field.  The classical expression for the area of ${\cal
S}$ is
\beq 
A({\cal S}) = 
\int_{\cal S}  d^2u\ \sqrt{\det g_{mn}(u)}, 
\eeq
where 
\beq 
g_{mn}(u) =\frac{\partial x^a(u)}{\partial u^m}
\frac{\partial x^b(u)}{\partial u^n} g_{ab}(x(u))
\eeq
is the two-dimensional metric induced on the surface.  The area
operator is constructed by expressing the area in terms of the
variable canonically conjugated to $A$, which is the inverse
densitized triad $E^a_i(x), i=1,2,3$, related to the metric by $\det
q\ q^{ab}=E_{i}^a(x)E_{j}^b(x)\delta^{ij}$.  This gives
\beq 
A({\cal S}) = \int_{A} d^2u\ \sqrt{n_a\ 
n_b\ E_{i}^{a}\ E_{j}^{b}\ \delta^{ij}}
\label{areac}
\eeq
where 
\beq n_{a}(u) = \epsilon_{abc}\frac{\partial x^b(u)}{\partial u^1}
\frac{\partial x^c(u)}{\partial u^2}.
\eeq
is the one-form normal to the surface ${\cal S}$.  If ${\cal S}$ is
singular at a point $p$, the normal $n_{a}(u)$ is not defined at $p$. 
This has no effect on the expression of the classical area
(\ref{areac}) because the singular point is a set of measure zero. 
Obviously, indeed, the area of a cone is defined in the same manner as
the area of a smooth surface.  However, what happens at single points
becomes important for the LQG quantum operator $A({\cal S})$ that
corresponds to the classical quantity (\ref{areac}).  A spin network
state determined by a spin network $S$ that crosses the surface at a
single point $p$ contributes to the area of the surface.  In the
derivation of this contribution, the tangent to the link of $S$ at $p$
gets contracted with the normal $n_{a}(p)$.  If this is ill defined,
we might expect a problem.  

The proper way of addressing this issue is in the context of a
quantization of the area operator based on a well-defined
regularization.  Several equivalent regularization schemes to define
area operator are discussed in the literature.  Not all of these
schemes can be immediately adapted to a surface with conical
singularities, but the regularization discussed in
\cite{Frittelli:1996cj}, which uses a smearing transversal to the
surface, remains well-defined for singular surfaces.  This
regularization is based on a continuous family of surfaces ${\cal
S}_{\lambda}$, with $\lambda \in [-\delta/2,\delta/2]$, where $\delta$
is a positive real number, such that ${\cal S}_{0}={\cal S}$. To
extend the technique to singular surfaces, we demand that
${\cal S}_{\lambda}$ is a smooth surface for $\lambda\neq 0$ and that
${\cal S}_{0}={\cal S}$ is singular.   The
area of $\cal S$ is then written as the limit
\beq 
A({\cal S}) = \lim_{\delta\to 0}\ \frac{1}{\delta}
\int_{-\delta/2}^{\delta/2} d\lambda\ A({\cal S}_\lambda) = 
\lim_{\delta\to 0}\ \frac{1}{\delta}
\int_{\mathcal{D}} d^3\sigma \sqrt{n_a n_b E^{a j} E^{b}_{j}}.
\eeq
where $\mathcal{D}={\cal S} \times [-\delta/2,\delta/2]$.  The non
vanishing contribution of the last integral comes now from the entire
one-dimensional intersection between the spin network and the three
dimensional region $\mathcal{D}$.  In this, the contribution of the
singular points of the $\lambda=0$ surface have measure zero.  The
fact that $\cal S$ has singular points is therefore irrelevant, and
the operator $A({\cal S})$ is well defined also for singular surfaces.

Recall that the LQG operator $A({\cal S})$ is ${\it Diff}$-covariant
in the sense that
\beq 
   A(\phi({\cal S})) = U_{\phi}\ A({\cal S}) \ U_{\phi^{-1}}
\label{AS}
\eeq
for all $\phi\in {\it Diff}$.  The above construction implies
immediately that (\ref{AS}) remains true also if $\phi\in {\it
Diff}^*$, because the differentiable structure at the intersection
point plays no role in the definition of $A({\cal S})$.  The action of
the area operator of a singular surface ${\cal S}$ is therefore
immediately obtained by (\ref{AS}) by choosing $\phi$ such that
$\phi({\cal S})$ is smooth.  We see that what matters is not the
linear structure at the intersection point, but just the topological
relation between the surface and the spin networks defining the
quantum states.

{} From this, it follows immediately that 
\begin{Pn}
The spectrum of the operator $A({\cal S})$ where ${\cal S}$ is
singular (has a finite number of conical singularities) is the same as
the spectrum of the operator $A({\cal S})$ where ${\cal S}$ is
smooth.
\end{Pn}
In conclusion, extended diffeomorphisms are indistinguishable from
ordinary diffeomorphisms as far as the area operator is concerned.  An
extended diffeomorphism may generate singular points in the surface or
in the spin network, but does not affect the topological relation
between a surface and the spin network, and the area depends only on
this relation.

\begin{figure}[t]
\begin{center}
\psfrag{a}{$\cal S$}
\psfrag{b}{$\phi({\cal S})$}
\psfrag{c}{$j_1$}
\psfrag{d}{$j_1$}
\psfrag{f}{$\phi$}
\includegraphics{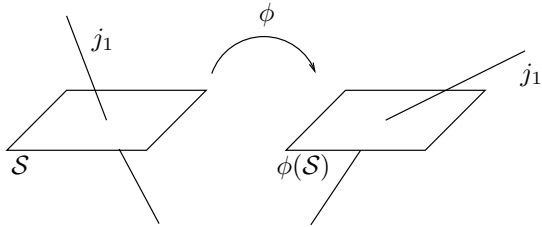}
\caption{An extended diffeomorphism may generate a conical singularity 
but 
does not change the topological relation between a surface $\cal S$ 
and the link of a spin network.}
\end{center}
\end{figure}

%%%%%%%%%%%%%%%%%%%%%%%%%%%%%%%%%%%%%%%%%%%%%%%%%%%%%%
\subsection{Volume and hamiltonian}
%%%%%%%%%%%%%%%%%%%%%%%%%%%%%%%%%%%%%%%%%%%%%%%%%%%%%%

Call $V({\cal R})$ the volume of a 3d region $\cal R$ in $\Sigma$.  There
exist two versions of the volume operator $V({\cal R})$ in LQG
\cite{Lewandowski}.  The first (let us call it $V_{1}({\cal R})$
here), used for instance in \cite{Rovelli:2004mc}, depends only on the
intertwiners of the nodes inside $\cal R$.  The second (let us call it
$V_{2}({\cal R})$ here), used for instance in \cite{Thiemann:2002nj},
depends also on whether or not the links at the nodes are linearly
dependent.  The operator $V_{1}({\cal R})$ is ${\it 
Diff}^*$-covariant, that is 
\beq 
   V_{1}(\phi({\cal R})) = U_{\phi}\ V_{1}({\cal R}) \ U_{\phi^{-1}}
\eeq
for all $\phi\in {\it Diff}^*$.  The operator $V_{2}({\cal R})$, on
the other hand, does not transform well under $\phi\in {\it Diff}^*$,
because an extended diffeomorphism can modify the linear dependence of
the links at the node.  Therefore the formulation of LQG considered
here requires the use of the version $V_{1}({\cal R})$ of the volume
operator.

Finally, the hamiltonian can be defined entirely in terms of the
volume operator and holonomy operators, and is not affected by the
modification of the theory considered here.

%%%%%%%%%%%%%%%%%%%%%%%%%%%%%%%%%%%%%%%%%%%%%%%%%%%%%%%%%%%%%%%%%%%%%
\section{Conclusion}
%%%%%%%%%%%%%%%%%%%%%%%%%%%%%%%%%%%%%%%%%%%%%%%%%%%%%%%%%%%%%%%%%%%%%

We have studied the problem of the separability of the
background-independent space of the quantum states of the
gravitational field, ${\cal H}_{\rm diff}$, in loop quantum gravity. 
We have shown that a small extension of the functional class of the
classical fields leads to an enlargement of the gauge group of the
theory.  In particular ${\it Diff}$ is enlarged to ${\it Diff}^*$, the
group of homeomorphisms that are smooth (with their inverse) except
possibly at a finite number of points.  The space of the knot classes
become countable and the kinematical Hilbert space ${\cal H}_{\rm
diff}$ is separable.  The area, volume, and hamiltonian operator are
naturally covariant under this extended gauge invariance, provided
that the appropriate regularization and the appropriate version of the
volume operator are chosen.  The spectra of area and volume, in
particular, are unaffected.  We expect that analogous results could be
obtained also using other mathematical settings, in particular the
piecewise smooth category.

We take these results as indications that the continuous moduli that
made ${\cal H}_{\rm diff}$ nonseparable might be physically spurious. 
Using the setting described in this paper, the theory appears to be
cleaner and to realize more completely its purely
combinatorial character as well as background independence.  If we
adopt this point of view, background independent quantum microphysics
is entirely discrete and smoothness can be seen, a posteriori, just as
a property arising from averaging over regions much larger than the
Planck scale.

\section*{\small Acknowledgments}

\indent Thanks to Philippe Roche for discussions and a careful reading
of the paper, to Alain Connes for captivating remarks on the
cross-ratio and its unfoldings and to Dan Christensen for clarification
on knot theory and for pointing out an imprecision in the first draft
of this paper.   

\section*{\small Note added in proofs}

After the posting of this work in the Archives, J. Lewandowski has
informed us that related ideas have been developed by him and A.
Ashtekar in work which is still unpublished.

\end{document}